\documentclass[sigconf,screen]{acmart}

\usepackage[most]{tcolorbox} 

\usepackage{adjustbox}
\usepackage{makecell}
\usepackage{multirow}
\usepackage{xspace}
\usepackage{xcolor}

\usepackage{soul}
\usepackage{hyperref}

\usepackage{tabularx}
\usepackage{array}
\usepackage{booktabs}
\usepackage{graphicx} 
\usepackage{caption}
\usepackage{threeparttable}
\usepackage{enumitem}

\newboolean{showcomments}
\setboolean{showcomments}{true}

\ifthenelse{\boolean{showcomments}}
{\newcommand{\nbc}[3]{
 {\colorbox{#3}{\bfseries\sffamily\scriptsize\textcolor{white}{#1}}}
 {\textcolor{#3}{\sf\small$\blacktriangleright$\textit{#2}$\blacktriangleleft$}}
 }
}
{\newcommand{\nbc}[3]{}
}

\newcommand\gopi[1]{\nbc{Gopi}{#1}{teal}}
\newcommand\keheliya[1]{\nbc{Keheliya}{#1}{purple}}

\newcommand{\AIBOM}{AIBOM}

\newcommand{\head}[1]{\par\smallskip\noindent\textbf{#1.}}

\definecolor{custom-gray}{cmyk}{0, 0, 0, 0.7}

\newtcbtheorem{Summary}{\bfseries Outcome}{enhanced,drop shadow={black!50!white},
  coltitle=black,
  top=0.15in,
  attach boxed title to top left=
  {xshift=1.5em,yshift=-\tcboxedtitleheight/2},
  boxed title style={size=small,colback=white}
}{summary}

\copyrightyear{2026}
\acmYear{2026}
\setcopyright{cc}
\setcctype{by-nc-nd}
\acmConference[ICSE-SEIP '26]{2026 IEEE/ACM 48th International Conference on Software Engineering}{April 12--18, 2026}{Rio de Janeiro, Brazil}
\acmBooktitle{2026 IEEE/ACM 48th International Conference on Software Engineering (ICSE-SEIP '26), April 12--18, 2026, Rio de Janeiro, Brazil}
\acmPrice{}
\acmDOI{10.1145/3786583.3786921}
\acmISBN{979-8-4007-2426-8/2026/04}
\begin{document}
\title{Building an Open AIBOM Standard in the Wild}
\subtitle{An Experience Report on Extending the SPDX SBOM (ISO/IEC 5962:2021) for AI Supply Chains}


\author{Gopi Krishnan Rajbahadur}
\email{grajbahadur@acm.org}
\orcid{0000-0003-1812-5365}
\affiliation{
  \institution{Queen's University}
  \city{}
  \country{Canada}
}

\author{Keheliya Gallaba}
\orcid{0000-0002-5880-5114}
\email{gallabak@sigsoft.org}
\affiliation{
  \institution{Queen's University}
  \city{}
  \country{Canada}
}

\author{Elyas Rashno}
\orcid{0000-0003-0813-3417}
\email{elyas.rashno@queensu.ca}
\affiliation{
  \institution{Queen's University}
  \city{}
  \country{Canada}
}

\author{Arthit Suriyawongkul}
\email{suriyawa@tcd.ie}
\orcid{0000-0002-9698-1899}
\affiliation{
  \institution{ADAPT Centre,\\Trinity College Dublin}
  \city{}
  \country{Ireland}
}
\author{Karen Bennet}
\email{karen.bennet@gmail.com}
\orcid{0009-0002-0353-3518}
\affiliation{
  \institution{IEEE}
  \city{}
  \country{USA}
}

\author{Kate Stewart}
\email{kstewart@linuxfoundation.org}
\orcid{0000-0003-1560-8329}
\affiliation{
  \institution{The Linux Foundation}
  \city{}
  \country{USA}
}

\author{Ahmed E. Hassan}
\email{ahmed@cs.queensu.ca}
\orcid{0000-0001-7749-5513}
\affiliation{
  \institution{Queen's University}
  \city{}
  \country{Canada}
}

\renewcommand{\shortauthors}{Rajbahadur et al.}


\begin{abstract}
Modern software engineering increasingly relies on open, community-driven standards, yet how such standards are created in fast-evolving domains like AI-powered systems remains underexplored. This paper presents a detailed experience report on the development of the AI Bill of Materials (\AIBOM{}) specification, an extension of the ISO/IEC 5962:2021 Software Package Data Exchange (SPDX) software bill of materials (SBOM) standard, which captures AI components such as datasets and iterative training artefacts. Framed through the lens of Action Research (AR), we document a global, multi-stakeholder effort involving over 90 contributors and structured AR cycles. The resulting specification was validated through four complementary approaches: alignment with major regulations and ethical standards (e.g., EU AI Act and IEEE 7000 standards), systematic mapping to six industry use cases, semi-structured practitioner interviews, and an industrial case study. Beyond delivering a validated artefact, our paper documents the process of building the \AIBOM{} specification ``in the wild,” and reflects on how it aligns with the AR cycle, and distills lessons that can inform future standardisation efforts in the software engineering community.
\end{abstract}



\maketitle
\section{Introduction}
\label{sec:introduction}

Standards play a critical role in Software Engineering (SE) by ensuring consistency, interoperability, reliability, and trustworthiness across systems and end users~\cite{fenton2002strategy}. They underpin much of the discipline. For instance, ISO/IEC 5962:2021, Software Package Data Exchange (SPDX), offers a shared vocabulary for documenting software supply chains (SSCs) that has been widely adopted to assess vulnerabilities and compliance in received software.

Regulators also rely on standards as mechanisms for compliance. In the EU, conformity with the Medical Device Regulation can be shown through harmonised standards such as ISO 13485 (quality management) and IEC 62304 (software lifecycle processes). In the U.S., Software Bill of Material (SBOM) standards like SPDX and CycloneDX are explicitly referenced in federal cybersecurity guidance \cite{sbommini2021, sbommini2025}. These SBOM standards have spurred a plethora of recent work that has examined SBOM adoption, proposed taxonomies, and developed extensions for vulnerability management \cite{Stalnaker2024, xia2024trust}, which underscores their growing importance.

These SBOM standards target traditional software and cannot meaningfully capture AI-specific artefacts such as datasets, trained models, fine-tuned checkpoints, and data pipelines that dominate modern systems~\cite{Xia2023}. To address this limitation, the AI community has introduced transparency frameworks such as model cards~\cite{Mitchell2019}, datasheets~\cite{gebru2021datasheets}, and factsheets~\cite{Arnold2019}, as well as machine-readable formats like Croissant~\cite{akhtar2024croissant}. Yet, these approaches often isolate AI-specific details from the broader software engineering context in which practitioners build and deploy systems. Sculley et al.~\cite{sculley2015hidden} showed in their seminal work on hidden technical debt that AI components typically form only a small fraction of much larger systems, underscoring the need to document them \textit{in situ} within the broader software context. Practitioners thus face a critical gap: they lack a unified, standard mechanism to describe AI components as first-class citizens within the software supply chain~\cite{rajbahadur2021can, hassan2024rethinking}.

\textbf{Our paper presents an experience report on tackling this challenge through the creation of the AI Bill of Materials (\AIBOM{}) specification, an extension to the SPDX 3.0 standard.}
Our extension comprises 36 new fields that treat datasets, models, and their provenance as first-class supply-chain elements to address the trustworthiness challenges. 
Our work makes a distinct contribution by showing how AR can be adapted from its traditional use within a single organisation~\cite{Baskerville1999} to guide a global, multi-stakeholder standardisation effort ``in the wild.” Through this lens, our \AIBOM{} specification itself emerges as a validated artefact of a disciplined, iterative process. Over 90 contributors participated in \textit{diagnose-design-evaluate-reflect} cycles, providing a replicable blueprint for developing standards in fast-moving domains in SE, while delivering a practical, community-driven specification.

\AIBOM{} specification has since landed in the SPDX standard, been piloted in industrial settings, and informed regulatory discussions. While the technical specification and schema live in our whitepaper~\cite{spdx3whitepaper},
this paper reports our \emph{experience} in open standardisation, detailing our governance, cadence, and pitfalls. We also demonstrate \AIBOM{} specification's practical utility using four complementary validation methods: (1) \textbf{regulatory/standards alignment} (EU AI Act, US and EU medical-device guidance, IEEE 7000 series) to support compliance practice; (2) \textbf{systematic mapping to industry use cases} (compliance, risk management, supply-chain governance) with coverage analysis; (3) \textbf{practitioner interviews} (n=10) across roles confirming field relevance and surfacing gaps; and (4) a \textbf{multinational industrial case study} showing \AIBOM{} covers ethics/legal checklists, generates portions of model cards, and enables partial automation for third-party AI assets. Together, these results position \AIBOM{} specification as part of an actionable standard and our process as a blueprint for creating standards in fast-moving, multi-stakeholder SE domains.

\textbf{Contributions.} We report how we created \emph{\AIBOM{}} specification through an open, global AR process and why this approach worked in practice. Our aim is not to claim that SPDX \AIBOM{} specification is the definitive AIBOM solution as credible alternatives such as CycloneDX exist and continue to evolve. Specifically:
\begin{itemize}[leftmargin=*]
  \item \textbf{AR at scale.} We demonstrate how AR, traditionally used within single organisations~\cite{Baskerville1999}, can be adapted to guide a global, multi-stakeholder standardisation effort “in the wild.” Our account documents governance design, release cadence, decision traceability, and iterative cycles that extended SPDX SBOM standard with 36 AI and Dataset specific fields.
  \item \textbf{Validated artefact.} We present \AIBOM{} and demonstrate its practicality via four complementary validations.
  \item \textbf{Lessons and blueprint.} We distill actionable patterns and lessons learned for creating standards in fast-moving, multi-stakeholder domains.
\end{itemize}

\section{Background}
\label{sec:background}





This section contextualises our work by examining the formal processes for creating software engineering standards, the technical foundation of the SBOM that requires extension.


\subsection{Development and Extension of Software Standards}

\smallskip\noindent\textbf{Standards in SE.}
Standards are vital in SE, providing common processes and terminology that minimise ambiguity, enhance interoperability, and ensure reliable high-quality system delivery~\cite{tsui2022essentials}. Current standards span lifecycle processes, quality assessment, testing, supply-chain security, and emerging AI considerations (Table~\ref{tab:se-standards}). These frameworks underpin the structured development and governance of modern enterprise software, fostering consistent engineering practices across diverse domains and organisations.

\begin{table}[t]
\centering
\scriptsize
\caption{Representative categories of software engineering standards and their scope.}
\begin{tabular}{@{}p{2cm}p{2cm}p{4cm}@{}}
\toprule
\textbf{Category} & \textbf{Examples} & \textbf{Purpose / Scope} \\
\midrule
Lifecycle process & ISO/IEC/IEEE 12207, 15288 & Define processes for development, maintenance, and complete system lifecycles. \\
Quality & ISO/IEC 25010 & Provide models for evaluating core software quality attributes. \\
Testing \& verification & ISO/IEC/IEEE 29119 & Standardise testing methods, documentation, and reporting practices. \\
Supply chain \& compliance & ISO 28000, GS1, CIS Controls & Secure supply chains and ensure regulatory and ethical conformance. \\
Emerging AI-related & ISO/IEC JTC 1/SC 42 & Address risk, transparency, and trust in AI systems. \\
\bottomrule
\end{tabular}
\
\label{tab:se-standards}
\end{table}

\smallskip\noindent\textbf{Standards development and extension pathways.}
Standards in SE typically evolve through two complementary pathways. The first is the \textbf{formal route}, followed by standards development organisations (SDOs) such as IEEE and ISO/IEC, which emphasise stability, consensus, and global legitimacy. This approach is characterised by well-defined, gated stages including project authorisation, working group (WG) formation, draft development, public review, and final balloting~\cite{iso2023}. These stages typically span 18--48~months for IEEE standards and 24--36~months for ISO/IEC standards~\cite{tsui2022essentials,ieeesa2023,iso2023}. The first version of IEEE POSIX standard, for example, progressed through this process via multiple iterations and ballots before becoming ISO/IEC~9945~\cite{isaak2005posix}.  

The second pathway is an \textbf{upstream, community-led extension model}, in which specifications evolve collaboratively in open WGs, consortia, or technical bodies before entering the formal process. This ``implementation-first'' or ``parallel'' model~\cite{blind2019relationship} underpins the evolution of major standards: POSIX is maintained by the Austin Group prior to IEEE and ISO ratification~\cite{austingroup2022}; amendments to IEEE~802.11 (Wi-Fi) are incubated in dedicated task groups~\cite{ieee80211process}; and SPDX itself matured within the Linux Foundation community before being standardised as ISO/IEC~5962:2021~\cite{spdxiso}. Such upstream-first approaches offer advantages such as faster iteration, broader participation, and validation against real-world practice. These factors are critical in fast-evolving domains such as AI. Our work deliberately adopts this second model.

\subsection{Software Bill of Materials (SBOMs)}

\smallskip\noindent\textbf{From BOM to SBOM.} The concept of a Bill of Materials (BOM) originated in manufacturing as a structured inventory of components and sub-assemblies within a product~\cite{back_8_jiao2000generic}. Applying this principle to software, a Software Bill of Materials (SBOM) provides a machine-readable inventory of software components, enabling organisations to understand what software they run, where it originates, and how it can be trusted~\cite{back_2_xia2023empirical}. SBOMs have become central to software supply chain security as modern systems increasingly depend on third-party and open-source components~\cite{Stalnaker2024}.

\smallskip\noindent\textbf{SBOM standards.}
Three SBOM standards dominate the landscape: SPDX~\cite{SPDX_linuxfoundation2025spdx}, developed under the Linux Foundation and standardised as ISO/IEC~5962:2021~\cite{spdxiso}; CycloneDX~\cite{cyclonedx_owasp2025cyclonedx}, introduced by OWASP and standardised as EMCA-424~\cite{ecma424}; and SWID Tags~\cite{SWID-iso2015swid}, maintained by the U.S. National Institute of Standards and Technology (NIST) to provide transparent software identification. While SPDX and CycloneDX are both widely adopted and offer comparable core capabilities, their selection in practice is typically guided by organisational priorities.



\smallskip\noindent\textbf{Limitations for AI systems.}
Despite their value, current SBOMs standards do not capture the full complexity of AI systems.
New artefact types, such as datasets and models, introduce deeper provenance, explainability, and compliance requirements. Dataset quality and lineage directly influence performance and fairness~\cite{back_26_lu2022towards,rajbahadur2021can}; model opacity necessitates metadata on interpretability, hyperparameters, and operational constraints~\cite{back_27_barclay2023providing}; and iterative lifecycles, including pre-training, fine-tuning, and redeployment, create intricate provenance chains and evolving compliance obligations~\cite{hassan2024rethinking}. These limitations have led both researchers and practitioners to call for extensions to SBOM standards that address the specific needs of AI systems, laying the foundation for the concept of an \AIBOM{}.

\section{Methodology}
\label{sec:methodology}

\begin{figure}[ht]
  \centering
  \includegraphics[width=\columnwidth]{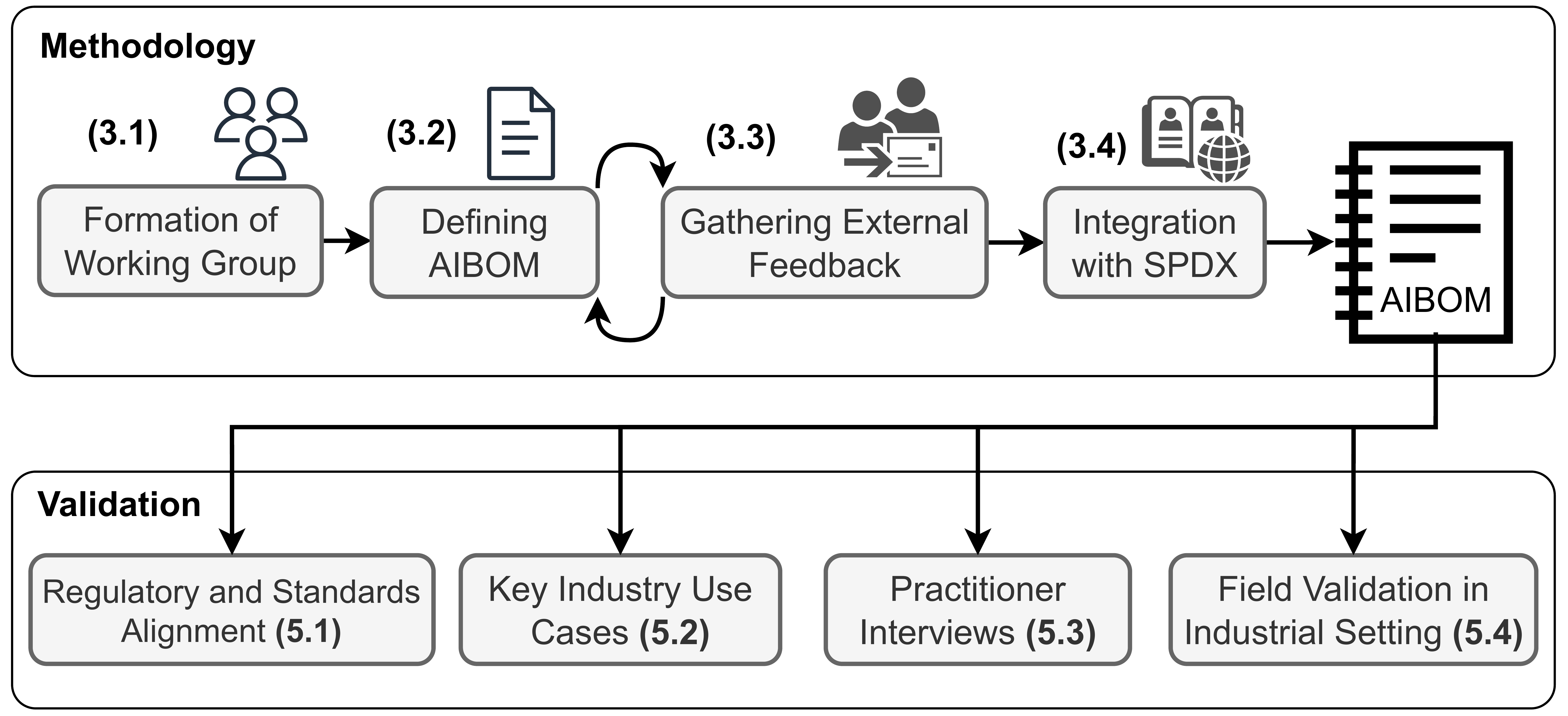}
  \caption{Overview of \AIBOM{} standard creation process and multi-faceted validation.}
  \label{fig:method-validation}
\end{figure}

We followed a participatory and iterative approach to develop the \AIBOM{} specification in the open, engaging multiple, diverse stakeholders. Figure~\ref{fig:method-validation} illustrates the process we followed.

\subsection{Working Group Formation}

\smallskip\noindent\textbf{Motivation and scope.}  
The AI Working Group (WG)~\cite{spdxaiwg} was formed in 2021 under the SPDX community to explore how existing SBOM standards could be extended to represent AI systems. Its primary goal was to determine whether SBOMs captured the artefacts and metadata needed for regulatory compliance, license obligations, auditability, and transparency in AI systems.

\smallskip\noindent\textbf{Recruitment and participation.}  
From the outset, the WG held weekly one-hour virtual meetings open to anyone interested. Early participation (2021 to mid-2022) relied on direct invitations and word-of-mouth recruitment through professional networks. To broaden engagement, the group launched a public mailing list on 7~June~2022, allowing contributors to sign up independently and join ongoing discussions.

\smallskip\noindent\textbf{Structure and operation.}  
The WG ran open agendas, made decisions by consensus, and used mailing lists and GitHub for asynchronous discussion and review. Participants included scientists, researchers, professors, CTOs, product managers, AI and software developers, and legal and licensing experts.

\begin{Summary}{}{xxx}
By May~2024, shortly after the official release of SPDX~3.0, the WG had held 82 meetings with 92 unique participants, 20 of whom attended more than six sessions. All meetings since April~2022 were recorded and archived publicly for transparency and traceability~\cite{spdxmeetingnotes}.
\end{Summary}

\subsection{Defining the \AIBOM{} Specification}

\smallskip\noindent\textbf{Initial field definition.}  
The WG’s first major task was to identify the core information an \AIBOM{} specification should capture. Drawing on their diverse expertise and relevant research, participants proposed an initial set of fields during weekly meetings in early 2022. The objective was to ensure that the \AIBOM{} could represent essential aspects of AI systems including datasets, models, and their associated metadata. At this stage, the focus was on brainstorming and cataloging all potentially relevant fields.

\smallskip\noindent\textbf{Incorporating existing practices.}  
From July~2022, once a comprehensive list had been drafted, members systematically analysed established documentation artefacts such as model cards~\cite{Mitchell2019}, datasheets~\cite{gebru2021datasheets}, and factsheets~\cite{Arnold2019} to refine and expand the field set. The goal was not exhaustive mapping, but to identify the most relevant and widely applicable fields recognised as useful and appropriate in the \AIBOM{} context. A detailed mapping of the fields from these sources that were incorporated, excluded and the reasons we did so, is provided in our whitepaper~\cite{spdx3whitepaper}.

\smallskip\noindent\textbf{Balancing adoption and completeness.}  
Throughout this process, the WG prioritised adoption over comprehensiveness by defining a minimal set of \emph{required fields}---those most likely to exist in real-world projects and satisfy regulatory or auditing needs. Each field had to meet strict inclusion criteria: (1) relevance to \AIBOM{} goals, (2) availability of a representation method, and (3) consensus on its necessity. More ambitious metadata elements were intentionally deferred to future releases. In line with NTIA~\cite{ntia2021sbom} recommendations and emerging best practices, the WG focused on a small, readily adoptable set of required fields. These fields were refined continuously until May~2024 through iterative review and discussion.

\begin{Summary}{}{THE}
The WG evaluated 103 candidate fields derived from existing tools and community proposals. The final specification defined 36 fields: 20 for the AI profile (five required) and 18 for the Dataset profile (six required).
\end{Summary}

\subsection{External Feedback and Public Consultation}
\label{sec:external_feedback}

\smallskip\noindent\textbf{Presenting drafts and gathering early feedback.}  
WG members frequently presented draft versions of the AI and Dataset profiles (that form the \AIBOM{} specification along with the rest of the SPDX profiles, please see Section~\ref{sec:aibom} for more details) at several formal venues, including Open Source Summit Europe (2022, 2023), Open Source Summit Japan (2022), and Open Source Summit North America (2023, 2024, 2025) and semi-formal WGs including the OpenChain Licensing WG~\cite{OpenChain_WG_Structure_2022}, OpenSSF AI WG~\cite{OpenSSF_AI_ML_Security_WG}, and IEEE P7014.1 WG~\cite{IEEE_P7014.1_WorkingGroup}. These events brought together open-source experts, software practitioners, and legal professionals, whose feedback was then discussed in WG meetings and incorporated into subsequent revisions of the specification.

\smallskip\noindent\textbf{Public release candidates and community review.}  
Starting on 8~May~2023, the WG initiated a year-long public consultation process, releasing two versions of the \AIBOM{} specification: two release candidates (RC1 and RC2) \cite{spdx_spdxspec_releases}. The SBOM tooling community was invited to review the specification, model, and profiles and to submit proposed changes as pull requests (PRs) to the public repository. In total, 20 PRs specific to \AIBOM{} were submitted, 16 of which were accepted after detailed discussion in WG meetings.

\smallskip\noindent\textbf{Integration with SPDX.}
In parallel with the release candidate process, the WG collaborated with the broader SPDX project to ensure that the proposed extensions were incorporated and could work seamlessly with other elements of the SPDX specification. Since several new fields required changes to the underlying data model, the group regularly presented proposed modifications to the SPDX Technical Team (responsible for the specification’s architecture and publication). These discussions led to coordinated updates across all SPDX WGs, enabling seamless integration of the \AIBOM{} related profiles and fields into SPDX~3.0 prior to its final release.

\begin{Summary}{}{Outcome} 
After multiple rounds of refinement, SPDX~3.0, including the AI and Dataset profiles (please see Section~\ref{sec:aibom} for more details) that encapsulate the \AIBOM{} specification was officially released on 16~April~2024. The release was accompanied by a Linux Foundation blog post~\cite{linuxfoundationSPDXRevolutionizes} and a detailed whitepaper~\cite{spdx3whitepaper} to support adoption and provide implementation guidance to the broader community.
\end{Summary}

\section{\AIBOM{}}
\label{sec:aibom}

\begin{figure}[t]
  \centering
  \includegraphics[width=\columnwidth]{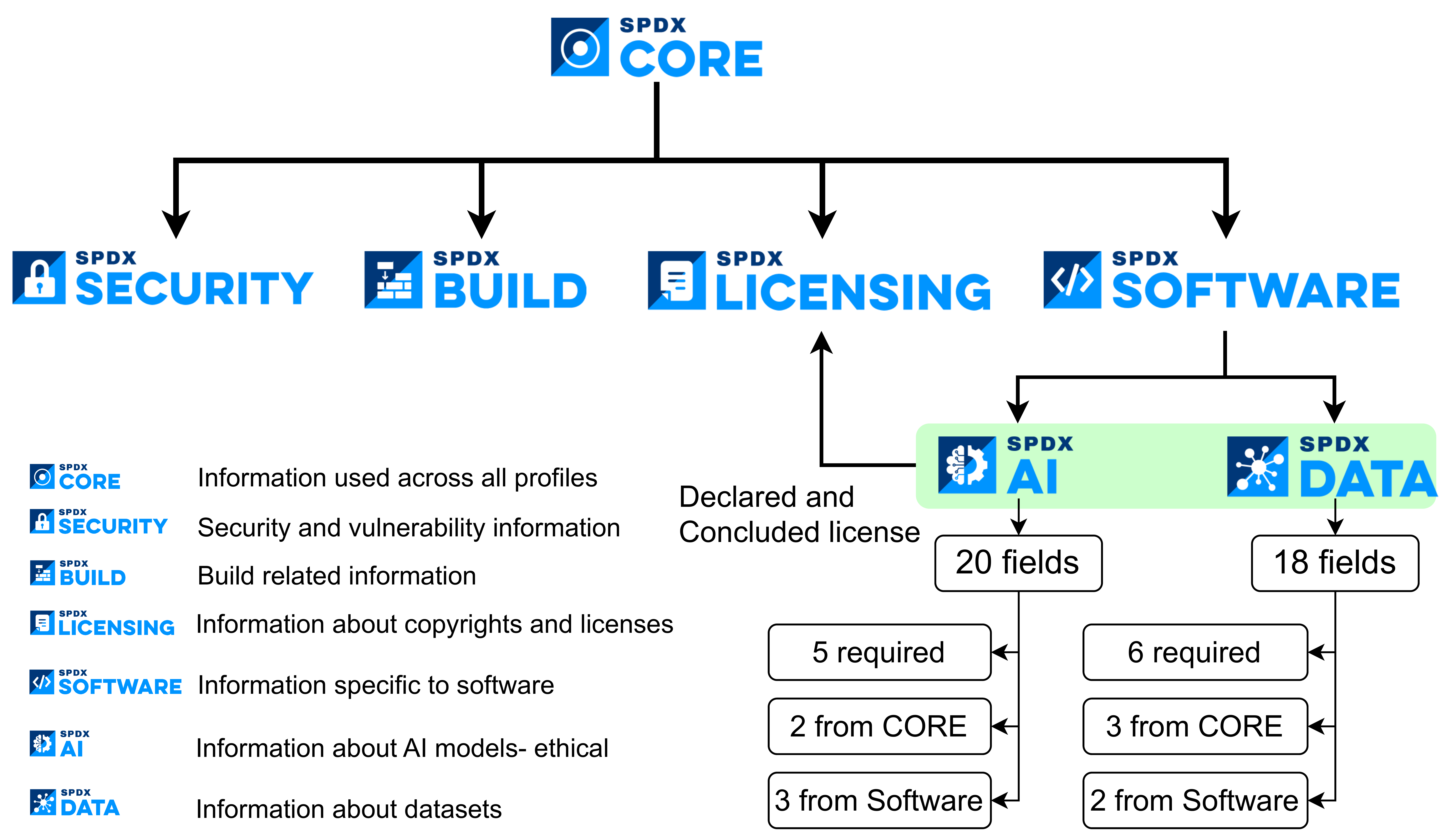}
  \Description{Diagram showing an overview of SPDX profiles and their extensions for AI and Dataset.}
  \caption{Overview of SPDX profiles and their extensions for AI and Dataset.}
  \label{fig:AIBOM}
\end{figure}

Building on the methodology described in Section~\ref{sec:methodology}, the working group integrated two new profiles: \textbf{Dataset} and \textbf{AI}, into the SPDX~3.0 specification. These extensions expand the SBOM standard beyond traditional software components to capture the artefacts underpinning AI systems. Figure~\ref{fig:AIBOM} shows how the various SPDX profiles together represent an \AIBOM{} specification (thereby extending the SBOM standard to represent AI systems. In our paper, we refer to this as the \AIBOM{} standard or \AIBOM{} specification. 

Designed to support regulatory compliance, licence attribution, provenance tracking, and responsible development, these extensions remain fully compatible with existing SPDX tooling. A detailed overview of the \AIBOM{} specification is available in the SPDX~3.0 white paper~\cite{spdx3whitepaper} and the public repository.\footnote{\url{https://github.com/spdx/spdx-3-model}} As this paper focuses on the specification's creation process over its technical details, we provide only a high-level summary below.

\smallskip\noindent\textbf{Profiles and extensibility in SPDX.}  
SPDX~3.0 adopts a profile-based modular architecture that enables domain-specific extensions while preserving interoperability. A \emph{profile} defines a coherent subset of the model: the \emph{Core} profile establishes foundational classes and vocabularies, while additional profiles extend these capabilities to represent licencing, build metadata, security, and software composition.

Within this framework, the \AIBOM{} specification for an AI system is realised through the \emph{AI} and \emph{Dataset} profiles working alongside existing components (though they can also be used independently to document models or datasets). Rather than exist as standalone artefacts, these profiles extend SPDX in a modular way without fragmenting the broader ecosystem (Figure~\ref{fig:AIBOM}). They build on existing classes and relationships from the \emph{Core}, \emph{Licensing}, and \emph{Software} profiles while introducing domain-specific metadata essential for describing AI artefacts. Together, they allow datasets and models to be represented as first-class supply-chain elements within the same SBOM framework.

\smallskip\noindent\textbf{AI profile.}
The AI profile captures AI model-specific components, capturing provenance (identifiers, versions, and licences), architecture (type, domain, and evaluation metrics), training details (data sources, configurations, and resource usage), and model risks (limitations, safety considerations, and compliance evidence). It provides a comprehensive view of how models are developed, deployed, and maintained.

\smallskip\noindent\textbf{Dataset profile.}
The Dataset profile standardises how datasets are described, capturing provenance (origins, sources, suppliers, and collection processes), descriptive details (size, modality, structure, and preprocessing), intended use (applications and purposes), and risk information (biases, sensitive data, and anonymisation methods). This enables traceability, accountability, and compliance throughout the AI lifecycle.

\section{Validation}
\label{sec:validation}

To demonstrate that our proposed \AIBOM{} specification is comprehensive, practical, and fit for purpose, we designed a multi-faceted validation strategy. This process evaluated our specification from four distinct but complementary angles which we detail in the following subsections.

\subsection{Regulatory and Standards Alignment}
\begin{table}[t]
\centering
\caption{Overview of key standards and regulations considered for \AIBOM{} alignment}
\small
\label{tab:standards-overview}
\scriptsize
\begin{tabular}{p{3cm} p{5cm} }
\toprule
\textbf{Studied Standard/Regulation} & \textbf{Objective of the Standard/Regulation}  \\
\midrule

\textbf{EU Artificial Intelligence Act (EU AI Act)}~\cite{europaRegulation20241689} & Ensure safe deployment, fundamental rights protection, and risk-based regulation of AI. \\

\textbf{EU Medical Device Regulation (EU MDR)}~\cite{RegulationEU20172017} & Demonstrate conformity and document risk management for medical devices placed on the EU market.  \\

\textbf{US Food and Drug Administration (FDA) Guidance}~\cite{fdaCodeFederal} & Support pre-market evaluation of software-based medical devices and promote cybersecurity risk management.  \\

\textbf{IEEE 7000 Series}~\cite{spiekermannIEEEP7000TheFirst2017} & Promote ethical design, transparency, bias mitigation, and data governance in autonomous and intelligent systems.  \\

\bottomrule
\end{tabular}
\end{table}

\smallskip\noindent\textbf{Motivation.}
It is essential that our \AIBOM{} specification captures all metadata required to verify compliance with key regulations and standards governing AI systems. Practical adoption depends on whether a supply chain standard enables Open Source Program Offices (OSPOs) and compliance auditors to fulfill their legal and governance obligations. In this section, we validate whether the fields defined in \AIBOM{} specification can represent the information demanded by widely adopted and emerging standards. We focus on the EU AI Act, U.S. and EU medical device regulations, and the IEEE~7000 series, as they collectively span regulatory, safety-critical, and ethical requirements across diverse domains. Table~\ref{tab:standards-overview} summarises the studied standards and regulations. Due to their extensive scope, we focused our analysis on the most relevant clauses and information elements that directly impact software supply chain compliance.

\smallskip\noindent\textbf{Methodology.}
We analysed the alignment between fields captured in the \AIBOM{} specification and the regulatory and ethical requirements defined by the standards and regulations shown in Table~\ref{tab:standards-overview}. The analysis was performed by the fourth and fifth author, who have extensive experience in standards development and regulatory alignment, and was further independently reviewed by first and the sixth authors for additional validation.

\smallskip\noindent\textit{EU AI Act.}, we examined the obligations defined in Articles~49 and~60, which govern mandatory registration in the EU high-risk AI database prior to market entry or testing in sandbox. Required information was grouped into four categories: \emph{Identification}, \emph{System Details}, \emph{Verification Details}, and \emph{Application Details}, comprising a total of 14 subcategories (e.g., provider contact, intended purpose, classification, testing plan, and involved parties).

\smallskip\noindent\textit{US and EU medical device regulations}, we analysed our \AIBOM{} specification's alignment with Article~10 of Regulation~(EU)~2017/745 and FDA cybersecurity guidance. Information elements were organised into three categories: \emph{package details}, \emph{model details}, and \emph{data details}.

\smallskip\noindent\textit{IEEE~7000 series}, we evaluated over 40 subclauses from eight standards (7000, 7001, 7002, 7005, 7007, 7009, 7010, and 7014), covering key areas such as transparency, privacy, data governance, and algorithmic bias.

\smallskip\noindent\textbf{Results.} 
Our analysis shows that the SPDX~3.0 \AIBOM{} specification provides strong coverage of regulatory, safety, and ethical requirements. We found that we could successfully represent \textbf{13 of 14} information obligations under Articles 49 and 60 of the EU AI Act, capturing \textbf{all required elements} from US and EU medical device regulations. Our validation shows that \AIBOM{} specification aligns with \textbf{over 40 subclauses} across eight IEEE~7000 series standards i.e., more than \textbf{90\%} overall coverage indicating practical readiness for compliance adoption.

However, the analysis also revealed limitations that highlight opportunities for future refinement. While \AIBOM{} captures most metadata required by the EU AI Act, it cannot yet explicitly represent the ``parties involved in testing” relationship mandated by Annex~IX~(2), even though contact details can be documented in existing fields. Similarly, the strong alignment with medical device regulations is partly due to their reliance on free-text “summary” and “description” fields, which map to generic SPDX constructs such as \texttt{informationAboutApplication} and \texttt{comment} but may limit traceability and automation. Finally, although more than 40 IEEE~7000 subclauses on transparency, privacy, data governance, and algorithmic bias are covered, areas such as child and student data governance, robotic nudging, and environmental and social governance remain outside the scope of the current \AIBOM{} specification. A detailed field-level mapping of these standards and their alignment with SPDX~3.0 is available in the official white paper~\cite{spdx3whitepaper}.

\subsection{Validation against Key Industry Use Cases}

\smallskip\noindent\textbf{Motivation.}  
While regulatory alignment ensures that our \AIBOM{} specification meets legal and compliance obligations, its practical value ultimately depends on how well it addresses real-world needs identified by industry stakeholders. In June~2025, the \textit{SBOM for AI Tiger Team} \cite{AIBOM_TigerTeam}, convened by the U.S. Cybersecurity and Infrastructure Security Agency (CISA), articulated six foundational use cases that any AI-focused SBOM standard must support. These use cases span the full lifecycle of AI system deployment: \textit{regulatory compliance}, \textit{Vulnerability and Incident Management},  \textit{legal and intellectual property assurance}, \textit{third-Party AI Risk Management}, \textit{Open Source Model Risk Assessment} and \textit{Model Lifecycle and Asset Management}. A detailed overview of these usecases can be found here~\cite{AISBOMTigerTeam}.  

Validating our \AIBOM{} specification against these use cases is therefore essential to demonstrate that the specification is not only standards-compliant but also operationally relevant, usable by organisations in practice, and capable of supporting the core security, legal, and governance workflows envisioned by the broader \AIBOM{} consumer ecosystem.

\smallskip\noindent\textbf{Methodology.}  
The CISA ``SBOM for AI" use cases are intentionally high-level; however, assessing whether our \AIBOM{} specification is capable of capturing the metadata outlined in them requires concrete, testable requirements. We therefore operationalised these use cases by consolidating them into broader themes, assembling a focused evidence base from recent studies, extracting atomic requirements, and mapping them to SPDX~3.0 \AIBOM{} fields before performing external validation.

\smallskip\noindent\textit{Step 1: Consolidate use cases into themes (with rationale).}  
Because several of the six use cases overlap conceptually and are instances of broader goals, we collapsed them into three themes to avoid double-counting and align with the structure of available evidence: (i) \textbf{Compliance} (Compliance; Legal \& IP Protections), (ii) \textbf{Open Source Risk Management} (Vulnerability/Incident Management; Open Source Model Risk Assessment), and (iii) \textbf{Supply Chain Management} (Third-Party AI Risk Management; Model Lifecycle \& Asset Management).

\smallskip\noindent\textit{Step 2: Build an evidence base per theme.}  
For each theme, we identified relevant research papers and technical reports that analyse concrete instances of the use cases. For example, Rajbahadur~et~al.~\cite{rajbahadur2021can}, which outlines methods for conducting dataset license compliance analysis, was mapped to the \textit{Compliance} category. All selected studies are listed in Table~\ref{table:AISBOM_UseCases}.

\smallskip\noindent\textit{Step 3: Extract atomic requirements.}  
The third author systematically reviewed each source and extracted concrete, verifiable requirements specific to its theme (template: \emph{requirement statement}, \emph{evidence snippet/page}, \emph{rationale}). For example, Rajbahadur~et~al.~\cite{rajbahadur2021can} examine licensing and provenance risks in using public datasets for commercial AI; from this, we derived the requirement to \emph{capture original data sources and licensing/redistribution constraints for each dataset and derivative build}. This process was repeated for all sources under each theme.

\smallskip\noindent\textit{Step 4: Map requirements to \AIBOM{}.}  
We created a traceability matrix linking each extracted requirement to specific SPDX~3.0 \AIBOM{} fields, indicating whether and how the requirement could be addressed by existing fields.

\smallskip\noindent\textit{Step 5: Internal and external validation.}  
The first author independently verified the extractions and mappings. We then presented the traceability matrix and representative examples to the \textit{AI SBOM Tiger Team} (29~Sept~2025, attended by 12 participants) and the \textit{SPDX AI and Dataset WG} (attended by four participants). Authors of this paper who were part of the working group did not participate in the validation exercise. Feedback was solicited on whether \AIBOM{} could be effectively operationalised to address the six use cases. Meeting minutes from both working group sessions are publicly available.\footnote{SBOM for AI (AIBOM) Tiger Team working group meeting minutes \url{https://docs.google.com/document/d/1IpXG7XBOJnPl_hwFf3JZkDaFb0k2CnI0/edit?usp=sharing&ouid=110194678381965933391&rtpof=true&sd=true}}

\smallskip\noindent\textbf{Results.}  
Our analysis demonstrated that the \AIBOM{} specification could represent all \textbf{46 distinct requirements} extracted from the eight studies reviewed across the three consolidated use-case categories. Table~\ref{table:AISBOM_UseCases} summarises the number of \AIBOM{} fields across the AI, Dataset, and other SPDX profiles that satisfied the requirements identified in each study. In several cases, a single requirement could be addressed by multiple fields, underscoring the flexibility and composability of our \AIBOM{} specification. A detailed traceability matrix, including the full set of extracted requirements and their field-level mappings, is available in our extended technical report~\cite{rashno2025aibom}.

\smallskip\noindent\textbf{External validation.}  
Feedback from participants in both the \textit{SPDX AI \& Datasets} and \textit{AI SBOM Tiger Team} working groups was broadly positive. Reviewers agreed that the methodology and resulting mappings were sound and well-aligned with practical needs, while noting that additional time would be required for an exhaustive, line-by-line review. As our primary objective was to identify any major conceptual gaps or misinterpretations rather than obtain final endorsement, this feedback was encouraging. Both groups have committed to conducting deeper evaluations in future iterations, and we intend to incorporate any outcomes available before the camera-ready version of this paper.

\begin{table}[t]
\centering
 

\scriptsize
\caption{Alignment of consolidated AI SBOM use cases and the number of mapped fields across the AI, Dataset, and other SPDX profiles.}
\begin{tabular}{lllll}
\toprule
\textbf{Categories} &
\makecell[l]{\textbf{References}} &
\makecell[l]{\textbf{AI} \\ \textbf{profile}} &
\makecell[l]{\textbf{Dataset}\\ \textbf{profile}} &
\makecell[l]{\textbf{Other} \\ \textbf{profiles}} \\
\midrule
\multirow{3}{*}{\makecell[l]{\textbf{Compliance}}} 
 & \cite{crfm2023} & 18 & 7 & 8 \\ [3pt]
 & \cite{rajbahadur2021can} & 5 & 11 & 3 \\ [3pt]
 & \cite{longpre2023data} & 4 & 3 & 3 \\ [3pt]

\midrule
\multirow{3}{*}{\makecell[l]{\textbf{Open Source Risk}\\\textbf{Management}}} 
 & \cite{thiel2023identifying} & 6 & 5 & 4 \\ [3pt]
 & \cite{jiang2024peatmoss} & 3 & 8 & 4 \\ [3pt]
 & \cite{stalnaker2025ml} & 5 & 4 & 2 \\ [3pt]

\midrule
\multirow{2}{*}{\makecell[l]{\textbf{Supply Chain}\\\textbf{Management}}} 
 & \cite{keskin2021cyber} & 7 & 3 & 3 \\ [3pt]
 & \cite{schlegel2023management} & 5 & 6 & 3 \\ [3pt]
\bottomrule
\end{tabular}
\label{table:AISBOM_UseCases}
\end{table}

\subsection{Validation based on Practitioner Interviews}
\smallskip\noindent\textbf{Motivation.}  
While the design of our \AIBOM{} specification was driven by a large, multi-stakeholder working group, it was essential to validate the resulting fields through an independent lens. To ensure that our specification was both understandable and practical beyond the community that developed it, we conducted semi-structured interviews with practitioners who were \emph{not} involved in the \AIBOM{} working groups. Their perspectives served as an external ``sanity check", allowing us to assess whether the proposed fields align with real-world needs, uncover overlooked requirements, and identify potential barriers to adoption in industry practice.

\smallskip\noindent\textbf{Methodology.}
This validation occurred prior to the official \AIBOM{} release, ensuring participants identified essential fields based on practical needs rather than reacting to the limitations of a finalised specification. Our methodology is detailed below.

\smallskip\noindent\textbf{Participant recruitment.}
We recruited ten practitioners via professional networks and targeted emails between 2022 and mid-2023. Participants were selected based on their expertise in SBOM development or AI software engineering to ensure a multi-disciplinary evaluation of the \AIBOM{} framework.

Table~\ref{table:study-participants} details the ten participants, who represent diverse work domains, experience levels, and potential future roles within the \AIBOM{} ecosystem.

\begin{table}[htbp]
  \centering
  \resizebox{\columnwidth}{!}{%
  \begin{threeparttable}
    \caption{Study Participant Demographics and Experience}
    \label{table:study-participants}
    \small 
    \setlength{\tabcolsep}{4pt} 

    \begin{tabular}{lrll}
      \toprule
\textbf{Part. ID} & \textbf{\makecell{Software/AI Dev./SBOM\\Experience (yrs)}} & \textbf{Domain} & \textbf{\makecell{\AIBOM{}\\Role}} \\
      \midrule
      P1\tnote{a}  & 11  / 9  / 0   & Entertainment Software & Consumer \\
      P2\tnote{b}  & 25  / 0  / 4   & Healthcare Software    & Consumer \\
      P3\tnote{c}  & 4   / 2  / 1   & ICT                    & Consumer \\
      P4\tnote{c}  & 14  / 2  / 2   & ICT                    & Both     \\
      P5\tnote{d}  & 15  / 6  / 0.5 & Academia               & Consumer \\
      P6\tnote{e}  & 20 / 0  / 16  & Open Source Foundation & Both     \\
      P7\tnote{f}  & 4   / 3  / 0   & Logistics Software     & Creator  \\
      P8\tnote{g}  & 12  / 12 / 0   & IT Consultancy         & Both     \\
      P9\tnote{d}  & 9   / 5  / 0   & Software Research      & Both     \\
      P10\tnote{h} & 15 / 0  / 5   & IT and Networking      & Both     \\
      \bottomrule
    \end{tabular}

 \begin{tablenotes}[para,flushleft] 
      \item[a] Lead Data Scientist
      \item[b] Chief Operating Officer
      \item[c] SE Researcher
      \item[d] Research Scientist
      \item[e] General Manager
      \item[f] ML Engineer
      \item[g] Consultant
      \item[h] SBOM Advocate
    \end{tablenotes}
  \end{threeparttable}
  }
\end{table}



\noindent\textbf{Interview protocol.}
After refining an initial questionnaire based on feedback from two academic and industry experts, we finalised an 18-question protocol. Interviews explored four themes: (1) participants' experience in AI and software development; (2) working definitions of AI software/system traceability; (3) current traceability practices and tools; and (4) expectations for the new \AIBOM{} specification.

We conducted semi-structured video interviews (30–60 minutes) to allow for improvisation and exploration~\cite{Wohlin2012}. To avoid bias, the proposed \AIBOM{} fields were not disclosed during the sessions. With participant consent, interviews were recorded, transcribed via software, and manually verified for accuracy.

\noindent\textbf{Transcript coding and analysis.} Two authors initially co-coded two transcripts to induce a baseline coding scheme. The remaining transcripts were then coded independently. Following this, the authors met to filter irrelevant codes and reach a consensus on categories. Finally, we computed the overlap between these participant-derived concepts and our proposed \AIBOM{} specification fields.

\begin{table}[t]
\caption{\AIBOM{} fields validated by interviews.}
\scriptsize
\centering
\begin{tabular}{p{1.8cm} p{2.3cm} p{3.7cm}}
\toprule
\textbf{Category} & \textbf{Proposed Field} & \textbf{SPDX 3.0 AI/Dataset Fields} \\
\midrule
\multirow{8}{*}{Data} 
  & Lineage & originatedBy, dataCollectionProcess, datasetUpdateMechanism, downloadLocation \\
  & Provenance & downloadLocation, originatedBy \\
  & Metadata & spdxId, name, packageVersion, buildTime, releaseTime, primaryPurpose, datasetSize \\
  & Assumptions & datasetNoise \\
  & Preprocessing & dataPreprocessing, anonymizationMethodUsed \\
  & Licenses & hasConcludedLicense, hasDeclaredLicense \\
  & Intended use & intendedUse \\
  & Personal info & hasSensitivePersonalInformation \\
\midrule
\multirow{8}{*}{Models} 
  & Behaviour & metric, metricDecisionThreshold, safetyRiskAssessment, modelExplainability \\
  & Parameters & hyperparameter \\
  & Algorithms & informationAboutTraining, typeOfModel \\
  & Intended use & primaryPurpose, informationAboutApplication \\
  & Assumptions & informationAboutApplication, limitation \\
  & Biases & knownBias, limitation \\
  & Licenses & hasConcludedLicense, hasDeclaredLicense \\
  & Features & modelExplainability, domain, typeOfModel \\
\midrule
Code & License & hasConcludedLicense, hasDeclaredLicense \\
\midrule
\multirow{3}{*}{Environment} 
  & Hardware & sensor, energyConsumption, inferenceEnergyConsumption, trainingEnergyConsumption \\
  & Configuration & proposed in SPDX 3.1 \\
  & Deployment & proposed in SPDX 3.1 \\
\midrule
\multirow{2}{*}{Process} 
  & Data Collection & dataCollectionProcess, sensor \\
  & Training & informationAboutTraining, trainingEnergyConsumption, finetuningEnergyConsumption \\
\midrule
\multirow{4}{*}{Governance} 
  & Dependencies & contains, downloadLocation \\
  & Attribution & originatedBy, suppliedBy, name, spdxId \\
  & Ethical considerations & safetyRiskAssessment, knownBias, useSensitivePersonalInformation, hasSensitivePersonalInformation, confidentialityLevel, anonymizationMethodUsed, standardCompliance, intendedUse \\
  & Design decisions \& limitations & limitation, modelDataPreprocessing, dataPreprocessing, metricDecisionThreshold, typeOfModel, hyperparameter \\
\bottomrule
\end{tabular}

\label{table:interview-field-mapping}
\end{table}

\noindent\textbf{Results.}
Excluding \textit{Configuration} and \textit{Deployment} fields—reserved for future versions due to capture complexity—all participant-proposed fields map to the SPDX \AIBOM{} 3.0 specification. Table~\ref{table:interview-field-mapping} details this mapping between participant suggestions and the specification.

The primary theme emerging from the interviews was the widespread inadequacy of current traceability practices for AI systems, which participants described as ad hoc and insufficient for ensuring proper governance. This sentiment was powerfully articulated by an executive director (P2), who stated, \textit{"I think it's a general mystery across the industry... I don't think that [traceability] adequately exists within the AI community around algorithms and training sets to be able to demonstrate that traceability from an auditing standpoint."} This lack of a standardised, auditable record is exacerbated by real-world development pressures. A machine learning engineer (P7) from a startup described this vividly: \textit{"It's something that we lack and it's because... the general nature of a startup, we usually deprioritise documentation and it usually also ends with a lot of pain for us... because we don't have the documentation, I usually spend so much time trying to explain anyway."} 

Beyond identifying these gaps, participants also provided clear insights into effective traceability requirements. When asked what information they need when consuming a pre-trained model, a lead data scientist (P1) offered a detailed wishlist: \textit{"What do I look for? I look for license... support... When, which training data was used, what demographic was used, and what biases do they have? ...And these are stuff that you want captured in the document, because that's what I'm looking for [as a model consumer]... what's the [reported] accuracy? How did you test it?"} P1’s requirements—spanning licensing, provenance, and evaluation—directly align with the \AIBOM{} framework’s core categories, confirming that its design addresses practitioner needs and will guide the specification's evolution.


\subsection{Field Validation in Industrial Setting}~\label{sec:industry_validation}

We evaluated \AIBOM{}'s practical suitability through an external study at a large multinational software organisation. From June to September 2024, a multi-disciplinary team of security architects, data scientists, DevOps experts, and AI developers mapped the overlap between our specification fields and those required during the organisation’s AI system development.

\head{Methodology} The organisation already had a checklist for developers to evaluate how their AI software development affects the internal ethics policy. Moreover, the legal department had a checklist to assess the legal impact of each AI-related software release.
The researchers investigated how many of the fields in these checklists could be populated if the organisations already had \AIBOM{s} constructed as part of the AI software development process. Furthermore, they investigated whether our \AIBOM{} specification can be used to automatically generate model cards. Then, they investigated to what extent \AIBOM{} generation for third-party models can be automated using web scraping or external APIs, such as Hugging Face~\cite{huggingface}.

\head{Results} Validation revealed that \AIBOM{} fields cover all internal checklist requirements for developers, proving its practical comprehensiveness. They observed that 60\% of existing model card fields are extractable from the fields in the \AIBOM{} specification. Furthermore, external APIs and web scraping successfully populated 40\% of fields for third-party models, demonstrating that \AIBOM{} generation can be partially automated to reduce manual effort.

\section{Standards Development as Action Research}
\label{sec:ar-discussion}
A primary contribution of this work is demonstrating how the development of a complex, multi-stakeholder standard can be systematically understood through the lens of \textbf{Action Research (AR)}. By retrospectively framing our effort to extend SPDX and create an \AIBOM{} specification as an AR process, we illustrate how community-driven standards development can maintain a disciplined, research-grounded approach within rapidly evolving domains.

While not initially conceived as AR, this project’s conduct and outcomes align closely with its core principles. AR focuses on changing practice through real-world intervention~\cite{Baskerville1999, Elden1993, Ferrario2014}; similarly, our objective was to develop a functional, community-driven standard for AI transparency and compliance. We therefore adopt the canonical AR cycle~\cite{staron2020} to retrospectively structure and reflect on this process.

Our AR framing is particularly relevant in software engineering (SE), where controlled experiments, case studies, and surveys dominate, while AR remains comparatively rare~\cite{petersen2014action}. Sjøberg et al.~\cite{Sjoberg2007} attribute this rarity partly to a limited understanding of AR’s role in SE. By analysing the \AIBOM{} specification development via an AR lens, we contribute a concrete example of its application in the wild to the software engineering body of knowledge.

Table~\ref{tab:ar_mapping} maps the main steps of our methodology to the canonical AR stages. This perspective highlights the key enablers of our success: continuous stakeholder engagement, iterative adaptation, and feedback-driven evolution. Further, it demonstrates how standards development can operate simultaneously as a scientific method and a practical intervention.

\begin{table}[t]
\centering
\caption{Mapping \AIBOM{} Development to the AR Cycle}
\label{tab:ar_mapping}
\small
\begin{tabularx}{\columnwidth}{p{0.32\columnwidth} X}
\toprule
\textbf{AR Phase} & \textbf{AIBOM Development Stage} \\
\midrule

\textbf{Diagnosing} \par \textit{Identifying a practical problem} &
\textbf{WG formation and scoping:} Addressed the lack of a standardised way to describe AI components for traceability, compliance, and transparency. \\

\textbf{Action Planning} \par \textit{Designing an intervention} &
\textbf{Field definition and framework design:} Extended SPDX to include AI artefacts by analysing model cards, datasheets, and existing practices. \\

\textbf{Action Taking} \par \textit{Implementing the intervention} &
\textbf{Drafting and presenting profiles:} Created draft AI and Dataset profiles and presented them at major forums to gather feedback. \\

\textbf{Observation} \par \textit{Evaluating impact} &
\textbf{Feedback collection:} Collected input via release candidates, pull requests, and discussions to assess utility and design. \\

\textbf{Reflection} \par \textit{Learning and refining} &
\textbf{Profile refinement:} Iteratively updated profiles based on feedback, aligning them with real-world adoption needs. \\

\textbf{Iterative Cycles} \par \textit{Repeating and improving} &
\textbf{Release iterations:} Integrated feedback into successive release candidates, refining the specification before final release. \\

\textbf{Knowledge Dissemination} \par \textit{Sharing outcomes} &
\textbf{Publishing and outreach:} Released SPDX~3.0 with AI and Dataset profiles, supported by a whitepaper, blog posts, and documentation. \\

\bottomrule
\end{tabularx}
\end{table}

\section{Lessons Learned and the Road Ahead}
\label{sec:discussion}

This section distills lessons from developing the \AIBOM{} standard in a multi-stakeholder setting and discusses how these insights inform future extensions for foundation model-powered software.

\subsection{Lessons Learned}~\label{sec:lessons}

\noindent\textbf{Present work-in-progress early and solicit feedback often.}
Public presentations of early drafts, in venues such as Open Source Summit (OSS) sessions (see Section~\ref{sec:external_feedback}) and other WGs provided critical feedback between mid-2022 and 2024. This early engagement not only strengthened the evolving standard but also built practitioner trust and buy-in. Feedback from outside our WG often prompted significant design changes. For instance, input from the OpenChain WG led to the introduction of a \texttt{standardsCompliance} field to capture standards an AI model or dataset already adheres to, which is a key requirement for audits. A question during an OSS North America 2023 panel about the queryability of fields such as \texttt{energyConsumption} and \texttt{metrics} prompted us to replace free-form text with structured types. Similarly, discussions at OSS Japan 2022 revealed that tracking AI and data lineage separately would complicate adoption, leading us to merge them into a unified \texttt{dataCollectionProcess} field. Based on our experience, in line with previous studies~\cite{eujrc}, we suggest that continuous feedback-driven development is not only advantageous but essential for open standards in fast-moving domains like AI.

\smallskip\noindent\textbf{Prioritise adoption over comprehensiveness.}~\label{sec:adoption}
Early drafts that attempted to capture every conceivable detail of an AI system consistently faced pushback from practitioners. Most organisations simply do not maintain information at that level of granularity, and \textit{a standard that demands it becomes impractical}~\cite{jespersen1995posix, isaak2005posix}. A similar pattern is evident with model cards, which prescribe numerous fields; over 60\% of models and 70\% of datasets on Hugging Face lack a complete model card~\cite{stalnaker2025ml,oreamuno2024state}. Several prior studies document similar resistance to heavyweight standards~\cite{pfleeger2002evaluating,schmidt2000implementing}. Another responsible AI field study reports an organisation expressing ``a real fear that they may adopt principles that they cannot uphold in practice''~\cite{ruster2025gaps}.

We therefore optimised our \AIBOM{} specification for adoption by defining a small set of readily recordable \emph{required fields} and enforcing strict entry criteria. In some cases, we intentionally excluded ambitious goals to improve practicality. For example, rather than attempting to enumerate every potential form of bias, we introduced a single \texttt{knownBias} field to capture only documented biases. Additional fields were deferred to future releases, enabling incremental evolution as practices mature.

We also found that two factors: metadata scale and availability, strongly influence adoption. For instance, a comprehensive \AIBOM{} for a self-driving car could span several gigabytes, making generation, storage, and maintenance impractical. At the same time, much of the desired metadata simply does not exist in accessible form. Even foundational datasets such as ImageNet and CIFAR-10 do not fully disclose their data sources~\cite{rajbahadur2021can}. Supporting fields that can be automatically collected or derived are therefore essential. Recognising this, we are now collaborating with the SPDX Implementers WG~\cite{SPDX_Implementers_List} and CISA SBOM-o-RAMA~\cite{CISA_SBOMaRama} to improve automation support in future iterations. These experiences highlight a broader reality: premature attempts at exhaustive coverage can undermine adoption. While similar lessons have been echoed several decades ago during POSIX standardisation~\cite{jespersen1995posix,isaak2005posix}, they ring true even today. Prioritising a minimal, usable core builds early momentum and a foundation for richer requirements as the ecosystem matures.

\smallskip\noindent\textbf{Plan for churn and conflict from day one.}
In long-running, community-driven standardisation, contributor turnover and conflicting priorities are inevitable; planning for them early is essential. Several participants disengaged as organisational priorities shifted, while others joined late to contribute to specific phases. This irregular participation often left critical tasks unfinished, hindering progress. Consequently, we established a rotating \emph{core group} with decision-making authority to ensure continuity alongside broader community input.

Disagreements were a constant reality in this multi-stakeholder setting. Participants frequently proposed modifications with vague or narrow motivations, risking developmental stalls. To mitigate this, we applied the evidence-based acceptance criteria defined in Section~\ref{sec:adoption}, significantly reducing friction and maintaining focus on shared priorities. As the AI ecosystem diversifies, sustaining continuity amid organisational churn and aligning competing interests will grow more challenging. Embedding these governance mechanisms early ensures open standards remain stable yet adaptable.

\subsection{Road Ahead}

\noindent\textbf{Evolving \AIBOM{} for FM-powered software.}
The 2023 surge in foundation models (FMs) and FM-powered software (FMware) necessitated an evolution of the AIBOM specification. Unlike static ML artefacts, FMware is continuous, adaptive, and open-ended~\cite{hassan2024rethinking}, introducing new assets like prompts, fine-tuned variants, and agents—the latter acting as first-class entities in multi-model pipelines~\cite{hassan2024rethinking, rajbahadur2024cool}. Operational layers such as \emph{grounding} (decision source verification) and \emph{guarding} (safety/policy enforcement) generate further dependencies requiring tracking~\cite{rajbahadur2024cool}. Driven by regulatory and risk-management needs, the FMwareBOM iteration expands \AIBOM{} to capture agent behaviours, orchestration context, and lifecycle transformations. While the current specification handles traditional AI (see Section~\ref{sec:validation}), adapting it for FMware is now urgent.

To meet FMware's requirements, we are updating the AI and Dataset profiles for the SPDX 3.1 release candidate~\cite{SPDX_spdx3_model_Milestone3}. This update introduces fields for agent identity and capabilities~\cite{spdx-1091}, enhanced provenance linking datasets to prompts and retrieval contexts~\cite{spdx-892}, and mechanisms to track adaptation events and fine-tuning lineage~\cite{spdx-1100}. These enhancements ensure FMwareBOM captures the full AI lifecycle while remaining automatable, auditable, and compliant with evolving standards.

\smallskip
\noindent\textbf{Making the future \AIBOM{} tool-native and automatable.}
As Section~\ref{sec:lessons} notes, effective tooling is vital for adoption, particularly as FMware increases tracking scale and complexity. To address this, we are designing the evolved version of \AIBOM{} to be tooling-native from the outset. Our goal is to ensure that key fields can be extracted directly from development pipelines or inferred through lightweight analysis, while still meeting regulatory and risk-management requirements. Our hope is that this approach will enable greater automation throughout the FMware lifecycle---from generating BOMs and validating schema conformance during CI stages to assessing compliance and policy obligations automatically. These design choices aim to make \AIBOM{} not only richer in representation but also a practical, automatable, and auditable part of the FMware development process.

\smallskip
\noindent\textbf{Enabling interoperability across the AI supply-chain ecosystem.}
Our roadmap positions the next-generation SPDX \AIBOM{} as a connective layer across a fragmented ecosystem. Rather than evolving in isolation, SPDX aims to provide a unified integration point for complementary standards—including SLSA (build provenance)~\cite{SLSA}, OpenChain (organisational compliance)~\cite{OpenChain_Project}, Croissant (data provenance)~\cite{akhtar2024croissant}, and Coalition for Secure AI~\cite{CoSAI_SAIF_Risk_Assessment}—ensuring interoperability across the AI supply chain. To address this, we are actively working to align and interoperate with these initiatives. For example, we are exploring how FMwareBOM can serve as the canonical machine-readable artefact for AI compliance within the OpenChain AI WG and collaborating with OpenSSF and CISA to integrate provenance, verification, and security metadata. To this end, our Open Source Summit North America 2025 panel on harmonising SLSA and SPDX~\cite{OSSNA2025_Panel_StrengtheningSupplyChains} gathered community input on reducing metadata duplication and ensuring standard reinforcement. Concurrently, the formal SPDX 3.0 specification is being prepared to update the ISO/IEC 5962 standard. This standardisation process, involving Technical Committee reviews, is expected to accelerate global adoption and ecosystem maturity.

\section{Related Work}
\label{sec:related_work}


To contextualise our contribution, we present prior work on emerging approaches for trustworthy AI, the challenges of multi-stakeholder collaboration in software engineering, and the application of AR as an intervention methodology.

\smallskip\noindent\textbf{Emerging approaches for trustworthy AI.}
Various initiatives address AI trustworthiness through auditability and provenance, yet gaps remain. DataBOM~\cite{back_5_liu2024blockchain} focuses on dataset provenance but lacks model-level detail, while CycloneDX ML-BOM~\cite{CycloneDXMLBOM} extends SBOMs to ML without capturing full lifecycle complexities. Reporting frameworks like factsheets~\cite{Arnold2019}, model cards~\cite{back_37_ModelCards}, and datasheets~\cite{gebru2021datasheets,Mitchell2019} enhance documentation but lack standardisation and system-level coverage. AI Cards~\cite{golpayeganiAICardsApplied2024} offer rich risk data but lack tooling. Broader governance frameworks (NIST AI RMF~\cite{back_39_ai2023artificial}, Microsoft RAI~\cite{microsoftMicrosoftResponsibleAI2022}, and Croissant~\cite{akhtar2024croissant}) promote accountability but remain decoupled from software supply chain realities. Recently, TAIBOM~\cite{nquiringmindsltdTAIBOM2025} aims to provide formal component and dependency descriptions. Despite their contributions, these solutions remain fragmented, treating datasets or models in isolation or offering high-level frameworks detached from implementation. This piecemeal landscape lacks a machine-readable artifact providing an end-to-end view of AI system composition. \AIBOM{} addresses this gap by extending the dependency-aware structure of SBOMs to unify data, models, and lifecycle processes into a cohesive, auditable standard.

\smallskip\noindent\textbf{SBOM research in SE.}
SBOMs are widely used for vulnerability management, transparency, risk assessment, and supply chain integrity~\cite{back_28_qvarfordt2024sbom, back_31_martini2023transparency}. Empirical work shows that they reduce remediation times by enabling rapid dependency analysis~\cite{back_29_benedetti2024impact} and improve auditability through provenance and licensing metadata~\cite{back_31_martini2023transparency}. However, their impact depends heavily on tooling completeness and metadata quality~\cite{Stalnaker2024, Xia2023}. Researchers and practitioners alike have proposed extensions, such as blockchain-enabled SBOMs for tamper resistance~\cite{xia2024trust} and DataBOMs for capturing provenance in data pipelines~\cite{back_4_barclay2019towards, back_5_liu2024blockchain}. However, our study is the first document the process of creating \AIBOM{} specification in a global, multi-stake holder setting. 

\smallskip\noindent\textbf{Standardisation experience reports.}
SE standardisation reports often provide historical narratives rather than reproducible extension methodologies. POSIX retrospectives trace evolution from user groups to ISO ratification but lack detail on how extensions were conceived~\cite{isaak2005posix, jespersen1995posix}. Similarly, the Austin Group’s SD/6 details governance but lacks practitioner perspectives~\cite{austinSD6}, while SystemVerilog reports document IEEE transitions without codifying development methods~\cite{sutherland2002systemverilog}. We extend this literature by detailing the end-to-end development of an AI extension—capturing decision-making and validation—and mapping standards development as a replicable Action Research (AR) cycle.

\smallskip\noindent\textbf{Multi-stakeholder studies in SE.}
Software engineering's socio-technical nature requires balancing the conflicting goals of developers, managers, customers, and regulators. Research shows that analysing these multi-stakeholder dependencies improves requirements elicitation~\cite{davis1993software} and risk management~\cite{alqahtani2017risk}, particularly within large-scale agile environments~\cite{hoda2017agile}. Challenges like interest balancing and cross-organisational communication~\cite{ralph2016role} are magnified in decentralised standards development. Lacking a single organisational authority, these volunteer-driven efforts face unique consensus-building and governance hurdles. We therefore use open standard development as a case study to analyse high-complexity multi-stakeholder collaboration.

\smallskip\noindent\textbf{Action Research in SE.}
AR has emerged as a key methodology for industry–academia collaboration, enabling iterative refinement of tools in real-world settings~\cite{avison1999action, petersen2014action}. While recent guidelines~\cite{staron2025guidelines} and hybrid models like Design Science Action Research~\cite{castro2025dsar} have matured the field, AR remains largely limited to internal process improvement within single organisations~\cite{dittrich2024industrial}. We extend AR to open, community-governed initiatives, using it as a framework to coordinate global stakeholders around a shared standardisation goal.

\section{Conclusion}
\label{sec:conclusion}
Our experience establishes Action Research as a robust framework for standardisation in dynamic, multi-stakeholder domains. By structuring the development of \AIBOM{} specification around iterative diagnose-design-evaluate-reflect cycles, we showed how open collaboration, continuous feedback, and evidence-based iteration can transform fragmented community efforts into a coherent, widely applicable standard. The lessons distilled from this process provide a blueprint for future initiatives like our evolution of \AIBOM{} specification in SPDX 3.1. Looking ahead, we envision AR-driven approaches playing a central role in shaping agile standards that evolve alongside the technologies they govern.


\head{Acknowledgements} We thank Helen Oakley and Rhea Michael Anthony for the industrial validation in Section~\ref{sec:industry_validation}; Prof. Daniel M. German and Prof. Zhen Ming (Jack) Jiang for their mentoring; and the members of the SPDX AI and Dataset Working Group for their contributions and feedback. Arthit Suriyawongkul's contribution to the WG is made possible through the financial support of Taighde Éireann – Research Ireland under Grant number 18/CRT/6224 (Research Ireland Centre for Research Training in Digitally-Enhanced Reality (d-real)).

\head{Disclaimer} Generative AI tools were used for copy-editing and table formatting. All experiments, analysis, and writing were performed by the authors, who thoroughly reviewed the content. This complies with IEEE and ACM policies on AI use in publications.

\bibliographystyle{ACM-Reference-Format}
\balance
\bibliography{references}

\end{document}